\documentclass[%
    aps,
    prl,
    twocolumn,
    reprint,
    10pt,
    superscriptaddress,
    shortbibliography,
    amsfonts,
    amssymb,
    amsmath,
    preprintnumbers,
    noshowpacs,
    noshowkeyws,
    tightenlines,
    floats,
    notitlepage,
    final,
    letterpaper,
    oneside,
    ]{revtex4-2}
\pdfoutput=1
\usepackage[utf8]{inputenc}
\usepackage[T1]{fontenc}
\usepackage[british]{babel}
\usepackage[british,cleanlook,printdayon]{isodate}
\usepackage[Symbolsmallscale]{upgreek}
\usepackage{mathtools}
\usepackage{isomath}
\usepackage[protrusion,tracking,kerning,spacing,final,babel]{microtype}
\usepackage{physics}
\usepackage[dvipsnames]{xcolor}
\usepackage[final]{hyperref}
\hypersetup{%
    colorlinks=true,
    citecolor=MidnightBlue,
    linkcolor=MidnightBlue,
    urlcolor=MidnightBlue
    }%
\usepackage{cleveref}
\usepackage{wasysym}
\usepackage{anyfontsize}

\newcommand{\asc}{\textit{\ascnode}}

\begin{document}
\title{Bridging the {\Large$\upmu$}Hz gap in the gravitational-wave landscape with binary resonance}

\author{Diego~Blas}
\affiliation{Grup de F\'isica Te\`orica, Departament de F\'isica, Universitat Aut\`onoma de Barcelona, 08193 Bellaterra, Spain}
\affiliation{Institut de Fisica d’Altes Energies (IFAE), The Barcelona Institute of Science and Technology, Campus UAB, 08193 Bellaterra, Spain}
\affiliation{Theoretical Particle Physics and Cosmology Group, Physics Department, King's College London, University of London, Strand, London WC2R 2LS, United Kingdom}

\author{Alexander~C.~Jenkins}
\email{alex.jenkins@ucl.ac.uk}
\altaffiliation{Corresponding author.\\Present address: Department of Physics and Astronomy, University College London, London WC1E 6BT, United Kingdom}
\affiliation{Theoretical Particle Physics and Cosmology Group, Physics Department, King's College London, University of London, Strand, London WC2R 2LS, United Kingdom}

\date{\today}
\preprint{KCL-PH-TH/2021-33}

%%%%%%%%%%%%%%%%%%%%%%%%%%%%%%%%%%%%%%%%%%%%%%%%%%%%%%%%%%%%%%%%%%%%%%%%%%%%%%%
\begin{abstract}
    Gravitational-wave (GW) astronomy is transforming our understanding of the Universe by probing phenomena invisible to electromagnetic observatories.
    A comprehensive exploration of the GW frequency spectrum is essential to fully harness this potential.
    Remarkably, current methods have left the $\upmu$Hz frequency band almost untouched.
    Here we show that this $\upmu$Hz gap can be filled by searching for deviations in the orbits of binary systems caused by their resonant interaction with GWs.
    In particular, we show that laser ranging of the Moon and artificial satellites around the Earth, as well as timing of binary pulsars, may discover the first GW signals in this band, or otherwise set stringent new constraints.
    To illustrate the discovery potential of these binary resonance searches, we consider the GW signal from a cosmological first-order phase transition, showing that our methods will probe models of the early Universe that are inaccessible to any other near-future GW mission.
    We also discuss how our methods can shed light on the possible GW signal detected by NANOGrav, either constraining its spectral properties or even giving an independent confirmation.
    \begin{center}
    \includegraphics[width=9pt]{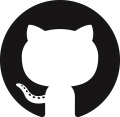}\hspace{2pt} Our results can be reproduced using the Python code \texttt{gwresonance}, available at \href{https://github.com/alex-c-jenkins/gw-resonance}{this URL}.
    \end{center}
\end{abstract}

\maketitle

%%%%%%%%%%%%%%%%%%%%%%%%%%%%%%%%%%%%%%%%%%%%%%%%%%%%%%%%%%%%%%%%%%%%%%%%%%%%%%%

\emph{Introduction.}---The direct detection of gravitational waves (GWs)~\cite{Abbott:2016blz} has initiated an exciting new era in astronomy, opening a window onto uncharted phenomena in the Universe.
The range of GW frequencies covered by current and future experiments will probe an impressive list of physical processes, from fundamental aspects of the early Universe to late-time astrophysical systems.
However, the practical limitations of these experiments leave certain windows in the GW spectrum unexplored.
Crucially, these windows may contain signals from new phenomena difficult to observe at other frequencies.
It is thus vitally important to cover the GW spectrum as thoroughly as possible.

A well-known gap in the GW landscape occurs at roughly $10^{-7}$--$10^{-4}$~Hz, between the sensitive bands of pulsar timing arrays (PTAs)~\cite{Lasky:2015lej,Arzoumanian:2020vkk,Janssen:2014dka} and future space-based interferometers such as LISA~\cite{Audley:2017drz}.
Accessing these frequencies is challenging, as this requires ``detectors'' of astronomical scale, which are nonetheless sensitive to the subtle effects of GWs.
One proposal is to construct a solar-system-sized interferometer~\cite{Sesana:2019vho}; however, such ideas remain futuristic.

Another possibility is to exploit the interaction of GWs with binary systems, an idea which has a long history~\cite{Bertotti:1973clo,Rudenko:1975tes,Turner:1979inf,Mashhoon:1981cos,Hui:2012yp}, but has yet to be fully explored.
Much like in any other system of masses, the passage of GWs through a binary perturbs the separation of the two bodies, leaving imprints on the system's orbit.
This effect is particularly pronounced if ($i$) the duration of the signal is much longer than the binary period, and ($ii$) the GW frequency is an integer multiple of the orbital frequency; the binary then responds resonantly to the GWs, allowing the perturbations to the orbit to accumulate over time.
By tracking changes in the binary's orbital parameters with sufficient precision, one can thus search for GWs at a discrete ``comb'' of frequencies set by the orbital period.
For periods ranging from days to years, this allows us to probe the $\upmu$Hz gap between LISA and PTAs.

We have recently developed a powerful formalism for calculating the evolution of a binary due to resonance with the \emph{stochastic GW background} (SGWB)~\cite{Blas:2021mpc}: the persistent, broadband signal sourced by the incoherent superposition of GWs from many sources that are too faint or too numerous to be resolved individually.
This formalism improves upon previous work~\cite{Bertotti:1973clo,Rudenko:1975tes,Turner:1979inf,Mashhoon:1981cos,Hui:2012yp} by capturing the evolution of the entire probability distribution for all six of the binary's orbital parameters.
In this Letter, we apply our formalism to explore the SGWB constraints that are possible with high-precision observations of various binary systems.
We show that Lunar laser ranging (LLR) and timing of binary pulsars can place stringent new bounds on the SGWB intensity in the $\upmu$Hz band, while satellite laser ranging (SLR) can be used to explore the LISA band in the decade before LISA flies.
Our forecast bounds span the entirety of the gap between LISA and PTAs, and are orders of magnitude stronger than all existing direct bounds in this frequency range.

We use units where $c=k_\mathrm{B}=1$, and set the Hubble constant to $H_0=67.66\,\mathrm{km}\,\mathrm{s}^{-1}\,\mathrm{Mpc}^{-1}$~\cite{Aghanim:2018eyx}.

%%%%%%%%%%%%%%%%%%%%%%%%%%%%%%%%%%%%%%%%%%%%%%%%%%%%%%%%%%%%%%%%%%%%%%%%%%%%%%%
\emph{Theoretical background.}---In the absence of perturbations, a Newtonian binary system traverses a fixed elliptical orbit, as determined by Kepler's laws.
This ellipse is described in terms of six orbital elements: $P$, the orbital period; $e$, the eccentricity; $I$, the inclination; $\asc$, the longitude of ascending node; $\omega$, the argument of pericentre; and $\varepsilon$, the mean anomaly at epoch.
If perturbed, for example by the passage of a GW, the binary will deviate from its Keplerian ellipse, causing its orbital elements to vary.
We thus treat these six parameters as functions of time, called the \emph{osculating} orbital elements~\cite{Murray:2000ssd,Blas:2021mpc}.

The SGWB is the most natural target of binary resonance searches, being persistent (rather than transient) and broadband (rather than narrowband).
The SGWB is also a highly interesting target, as it encodes the GW emission from a broad range of sources throughout cosmic history.
These sources are likely to include unresolved astrophysical systems at low redshift, such as inspiralling compact binaries~\cite{Regimbau:2011rp}, and may also include a host of more exotic early-Universe sources, including cosmological first-order phase transitions (FOPTs)~\cite{Caprini:2015zlo,Caprini:2019egz}, cosmic strings, and inflationary tensor modes~\cite{Caprini:2018mtu}.

The unpredictable arrival times and phases of GWs from many independent sources make the SGWB inherently random~\cite{Caprini:2018mtu}, and we therefore cannot hope to predict the exact evolution of the osculating elements for any given binary.
We can, however, calculate the statistical properties of this evolution, allowing us to predict the time evolution of the \emph{distribution function} (DF) of the orbital elements, $W(\vb*X,t)$, where $\vb*X=\{P,e,I,\asc,\omega,\varepsilon\}$.
This is defined such that an integral over any region $\mathcal{X}$ of parameter space gives the corresponding probability for the osculating elements taking those values at time $t$,
    \begin{equation}
        \Pr(\vb*X\in\mathcal{X}|t)=\int_\mathcal{X}\dd{\vb*X}W(\vb*X,t).
    \end{equation}
Assuming the SGWB perturbations are Gaussian, the time-evolution of the DF follows a nonlinear~\footnote{The FPE is clearly linear in $W$. In this context, ``nonlinear'' means that the drift vector and diffusion matrix are functions of the orbital elements, giving rise to interesting effects such as noise-induced drift~\cite{Risken:1989fpe}.} \emph{Fokker-Planck equation} (FPE)~\cite{Risken:1989fpe,Blas:2021mpc},
    \begin{equation}
    \label{eq:fpe}
        \pdv{W}{t}=-\pdv{X_i}(D_i^{(1)}W)+\pdv{X_i}\pdv{X_j}(D_{ij}^{(2)}W)\,,
    \end{equation}
    (with summation over repeated indices implied).
Here $D^{(1)}_i$ and $D^{(2)}_{ij}$ are the \emph{drift vector} and \emph{diffusion matrix}; functions of the orbital elements encoding the statistical properties of the stochastic perturbations.
In our case, these quantities are fully specified by the SGWB intensity spectrum,
    \begin{equation}
        \Omega_\mathrm{gw}(f)\equiv\frac{1}{\rho_\mathrm{c}}\dv{\rho_\mathrm{gw}}{(\ln f)},
    \end{equation}
    which is the energy density in GWs per logarithmic frequency bin, normalised relative to the critical energy density of the Universe, $\rho_\mathrm{c}\equiv3H_0^2/(8\uppi G)$.
In a companion paper~\cite{Blas:2021mpc} we derive $D^{(1)}_i$ and $D^{(2)}_{ij}$ for a binary immersed in a Gaussian SGWB; both can be written as linear combinations of the SGWB intensity at the binary's harmonic frequencies,
    \begin{align}
    \begin{split}
        D^{(1)}_i(\vb*X)&=V_i(\vb*X)+\sum_{n=1}^\infty\mathcal{A}_{n,i}(\vb*X)\Omega_\mathrm{gw}(n/P),\\
        D^{(2)}_{ij}(\vb*X)&=\sum_{n=1}^\infty\mathcal{B}_{n,ij}(\vb*X)\Omega_\mathrm{gw}(n/P).
    \end{split}
    \end{align}
Note that the drift vector also includes a deterministic term $V_i$ accounting for the binary's evolution in the absence of the SGWB.
This includes relativistic effects such as the precession of the pericentre $\omega$ and the decay of the period $P$ and eccentricity $e$ due to radiation of GWs, which are particularly important to capture in the case of binary pulsars.

To get a sense of how strong we can expect our forecast constraints to be, it is instructive to carry out a back-of-the-envelope calculation in which the rms perturbation to the orbital period after time $T$ is $\sigma_P=\sqrt{2TD^{(2)}_{PP}}$.
Taking the LLR case as an example, for a SGWB intensity $\Omega_\mathrm{gw}=10^{-5}$ and an observation period of $T=15\,\mathrm{yr}$, this gives $\sigma_P\sim1\,\upmu\mathrm{s}$.
This corresponds to a rms perturbation to the semi-major axis of $\sigma_a=(2a/3P)\sigma_P\sim0.1\,\mathrm{mm}$.
Given that each LLR ``normal point'' measurement determines the Earth-Moon distance to within $\sim3\,\mathrm{mm}$, we see that a campaign of $\sim1000$ such measurements should be capable of detecting this signal.

%%%%%%%%%%%%%%%%%%%%%%%%%%%%%%%%%%%%%%%%%%%%%%%%%%%%%%%%%%%%%%%%%%%%%%%%%%%%%%%
\emph{Results and discussion.}---Our main results are based on three different high-precision probes of binary orbital dynamics:
    \begin{description}
        \item[MSP] Timing of binary millisecond pulsars (MSPs), with periods between $P\approx1.5\,\mathrm{hr}$ and $P\approx5.3\,\mathrm{yr}$~\cite{Manchester:2004bp};
        \item[LLR] Laser-ranging measurements of the Moon's orbit around the Earth ($P\approx27\,\mathrm{days}$)~\cite{Murphy:2013qya};
        \item[SLR] Laser-ranging measurements of the orbits of artificial satellites around the Earth, in particular the LAGEOS-1 satellite ($P\approx3.8\,\mathrm{hr}$)~\cite{Ciufolini:2016ntr}, as this has been regularly producing laser-ranging data for longer than any other satellite mission.
    \end{description}

We numerically evolve the first and second moments of the FPE~\eqref{eq:fpe} from delta-function initial conditions for each of these systems using our Python code \href{https://github.com/alex-c-jenkins/gw-resonance}{\texttt{gwresonance}}, which we make publicly available at the linked URL.
This gives a probabilistic model for the orbital elements over time, which we combine with a Fisher-forecasting approach to calculate the expected sensitivity of each binary to the SGWB.
(See the Supplemental Material~\footnote{The Supplemental Material, which includes Refs.~\cite{Poisson:2014gr,Peters:1963ux,Blandford:1976pt,Casella:2002stat,Lyne:2015oua,Abdo:2010en,Kramer:2006nb,Weisberg:2016jye,Jacoby:2006dy,Fonseca:2014qla,Haniewicz:2020jro,vanKerkwijk:1999xj,Kaspi:1996pul,Freire:2010tf,Madsen:2012rs,Miller-Jones:2018waj,Liu:2011cka,ilrs:lageos,Fedderke:2020yfy,jpl:sbdb,Schmitz:2020syl,Espinosa:2010hh,Ellis:2020awk,Desjacques:2020fdi,Armaleo:2020yml,Blas:2019hxz}, provides further details about our procedure for integrating the Fokker-Planck equation and computing sensitivity forecasts, as well as details on the FOPT spectra we use to generate Fig.~\ref{fig:fopt}, so that all of our results may be reproduced using our publicly-available code \href{https://github.com/alex-c-jenkins/gw-resonance}{\texttt{gwresonance}}.
We also discuss how to obtain long-timescale SGWB constraints using the present-day orbital parameters of Solar System bodies.}.)

\begin{figure}[t!]
    \includegraphics[width=0.485\textwidth]{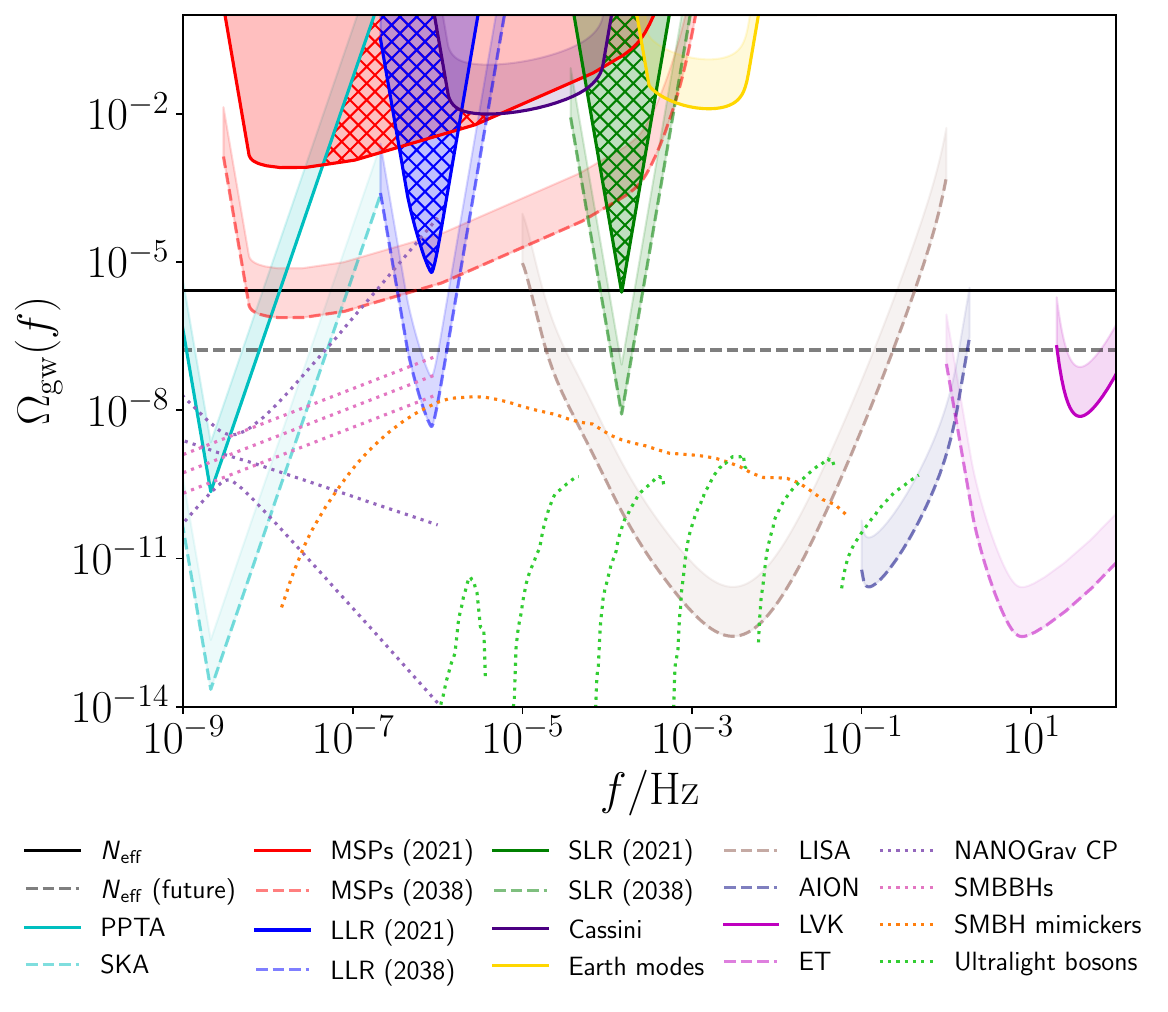}
    \caption{%
    SGWB sensitivity curves of current and future GW experiments, as well as our forecasts.
    Each curve is a 95\% confidence upper limit ($\mathrm{SNR}=2$), with shaded regions extending up to $\mathrm{SNR}=20$.
    Solid curves indicate existing results from the LIGO/Virgo/KAGRA Collaboration~\cite{KAGRA:2021kbb,Abbott:2021pi} (LVK), gravimeter monitoring of the Earth's normal modes~\cite{Coughlin:2014xua}, Doppler tracking of the Cassini spacecraft~\cite{Armstrong:2003ay}, pulsar timing by the Parkes PTA~\cite{Lasky:2015lej}, and indirect constraints from $N_\mathrm{eff}$~\cite{Pagano:2015hma}, as well as our forecast present-day sensitivities for binary resonance searches with binary millisecond pulsars (MSPs), Lunar laser ranging (LLR), and satellite laser ranging (SLR), which are presented for the first time here.
    Hatching indicates the new region probed by our present-day forecasts.
    Dashed curves indicate our binary resonance forecast sensitivities for 2038, along with expected bounds from ET~\cite{Punturo:2010zz}, LISA~\cite{Audley:2017drz}, SKA~\cite{Janssen:2014dka}, and the proposed km-scale atom interferometer AION~\cite{Badurina:2019hst}, as well as improved $N_\mathrm{eff}$ constraints~\cite{Pagano:2015hma}.
    Dotted curves show various potential SGWB signals in the $\upmu$Hz band.
    The purple curves indicate a possible signal associated with the common process (CP) identified by NANOGrav~\cite{Arzoumanian:2020vkk}, while the overlaid pink curves show the inferred amplitude for the NANOGrav CP when assuming a $\Omega_\mathrm{gw}\sim f^{2/3}$ spectrum, as expected for SMBBHs.
    The yellow curves show two FOPT spectra at temperatures $T_*=2\,\mathrm{GeV}$ and $200\,\mathrm{GeV}$, peaking at $f\approx1\,\upmu\mathrm{Hz}$ and $\approx100\,\upmu\mathrm{Hz}$ respectively.
    The orange curve shows the predicted spectrum from a population of horizonless SMBH mimickers~\cite{Barausse:2018vdb}.
    The pale green curves show the predicted spectra from ultralight bosonic condensates around SMBHs~\cite{Brito:2017wnc}, with boson masses varying from $10^{-20}\,\mathrm{ev}$ (left-most curve) to $10^{-15}\,\mathrm{eV}$ (right-most curve).}
    \label{fig:all-constraints}
\end{figure}

\begin{figure}[t!]
    \includegraphics[width=0.48\textwidth]{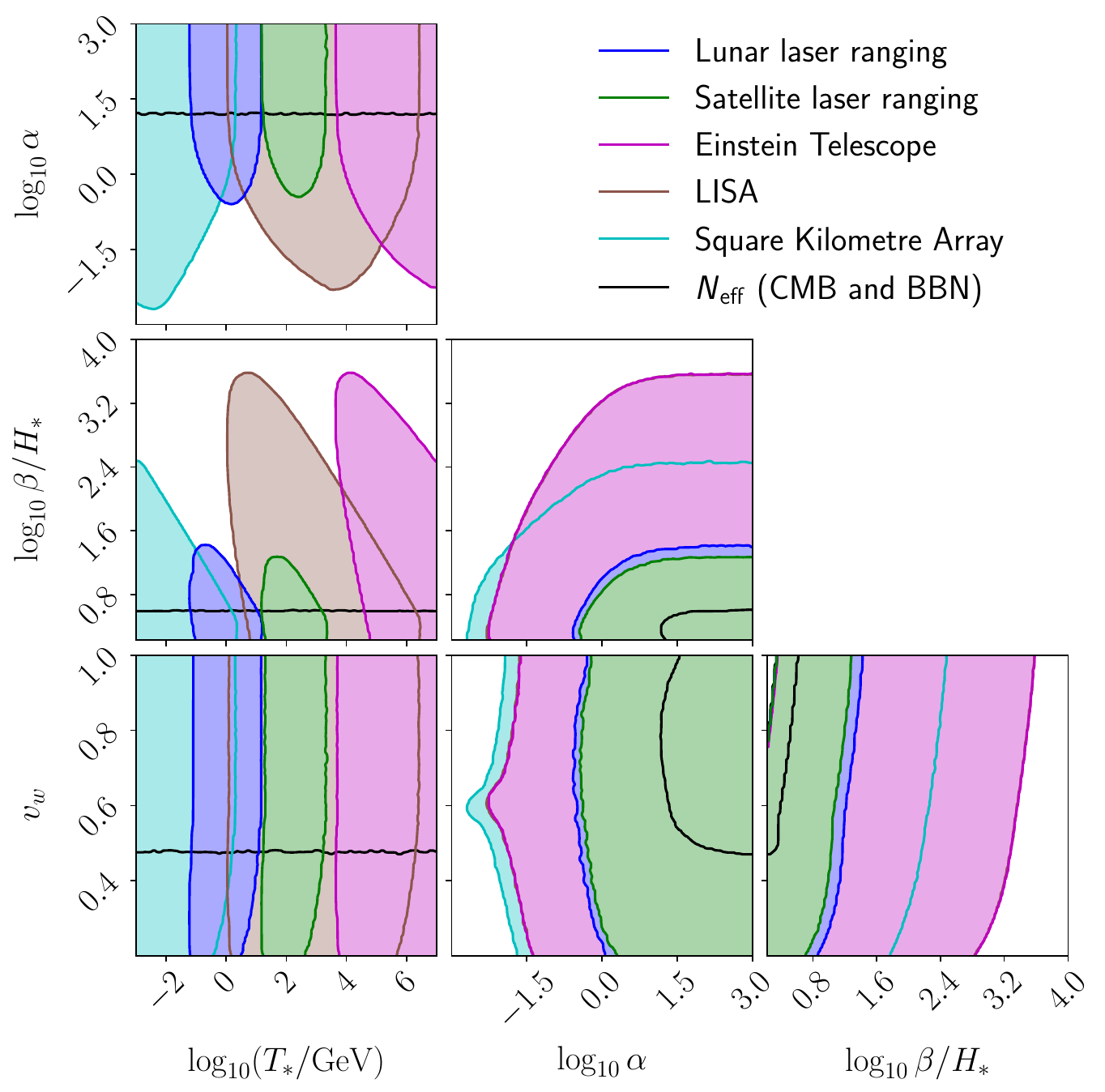}
    \caption{%
    Forecast exclusion regions of the FOPT parameter space for various SGWB searches at 2038 sensitivity.
    Here $T_*$ is the temperature at which the FOPT occurs, $\alpha$ is the energy density released by the FOPT in units of the radiation density at the transition epoch, $\beta/H_*$ is the inverse duration of the transition in units of the Hubble rate at the transition epoch, and $v_w$ is the bubble wall velocity.
    }
    \label{fig:fopt}
\end{figure}

The resulting power-law integrated (PI)~\cite{Thrane:2013oya} sensitivity curves are shown in Fig.~\ref{fig:all-constraints}, alongside the sensitivities of various other current and future GW experiments~\footnote{Note that these curves represent the sensitivity to a power-law SGWB, $\Omega_\mathrm{gw}(f)\sim f^\alpha$ with $|\alpha|\le10$, and that sufficiently sharply-peaked spectra might overlap these curves without being detected; see the Supplemental Material for further details.}.
For each of our binary resonance probes (MSP, LLR, and SLR), we calculate two sensitivity curves: one which reflects the data available in 2021, and one which should be achievable by 2038, by which time LISA is expected to have completed its nominal 4-year mission.
By this point in the late 2030s we also anticipate sensitive SGWB searches by the Einstein Telescope~\cite{Punturo:2010zz} (ET; a planned third-generation GW interferometer), the Square Kilometre Array~\cite{Janssen:2014dka} (SKA; a radio telescope array whose planned uses include a next-generation PTA to search for nHz GWs) and by some km-scale versions of the atom interferometers AION~\cite{Badurina:2019hst} or MAGIS~\cite{Abe:2021ksx}, which occupy the frequency band between LISA and ground-based interferometers.
(There are various other constraints at lower frequencies not shown here, including those from CMB temperature and polarisation anisotropies~\cite{Ade:2015tva,Namikawa:2019tax} and spectral distortions~\cite{Kite:2020uix}, as well as potential future constraints in the frequency band we are interested in, e.g. from astrometry~\cite{Moore:2017ity,Darling:2018hmc,Wang:2020pmf,Garcia-Bellido:2021zgu,Aoyama:2021xhj}, helioseismology~\cite{Lopes:2015pca}, modulation of GW signals~\cite{Bustamante-Rosell:2021daj}, the $\upmu$Ares proposal~\cite{Sesana:2019vho}, the Moon's normal modes~\cite{LGWA:2020mma,Jani:2020gnz}, and high-cadence PTA observations~\cite{Perera:2018pts,Wang:2020hfh}.
However, all these constraints are either very futuristic, not applicable to stochastic GW signals, or not strong enough to be competitive with our forecasts.)
The horizontal black lines in Fig.~\ref{fig:all-constraints} show indirect constraints due to SGWB contributions to the effective number of relativistic degrees of freedom ($N_\mathrm{eff}$) in the early Universe~\cite{Pagano:2015hma}, as probed by the Cosmic Microwave Background (CMB) and Big-Bang Nucleosynthesis (BBN).
These lines should be interpreted differently from the other constraints that we show, as they represent bounds on the \emph{total} sub-horizon SGWB energy density [the values plotted correspond to the upper bounds on $\int\dd{(\ln f)}\Omega_\mathrm{gw}$ at frequencies $f\gtrsim10^{-15}\,\mathrm{Hz}$], and only include GWs emitted before the epoch of BBN.

We find that laser-ranging experiments are already able to place cosmologically relevant bounds with present data; LLR has an expected sensitivity of $\Omega_\mathrm{gw}\ge6.2\times10^{-6}$ at $f=0.85\,\upmu\mathrm{Hz}$ ($95\%$ confidence upper limit), while the forecast for SLR with the LAGEOS satellite is $\Omega_\mathrm{gw}\ge2.4\times10^{-6}$ at $f=0.15\,\mathrm{mHz}$.
These forecasts, if realised, would be by far the most sensitive direct SGWB searches to date in the broad frequency band between ground-based interferometers at $f\gtrsim10\,\mathrm{Hz}$ and PTAs at $f\sim\mathrm{nHz}$, a full three orders of magnitude stronger than existing constraints from the Cassini spacecraft~\cite{Armstrong:2003ay} and the Earth's normal modes~\cite{Coughlin:2014xua}, and competitive with indirect $N_\mathrm{eff}$ constraints~\cite{Pagano:2015hma}, which currently set $\int\dd{(\ln f)}\Omega_\mathrm{gw}\le2.6\times10^{-6}$.
With some reasonable assumptions about future improvements in the noise levels and data cadence of laser-ranging experiments (see the Supplemental Material), these forecasts improve to $\Omega_\mathrm{gw}\ge4.8\times10^{-9}$ for LLR and $\Omega_\mathrm{gw}\ge8.3\times10^{-9}$ for SLR by 2038, significantly better than the $N_\mathrm{eff}$ constraint, which is expected to reach $\int\dd{(\ln f)}\Omega_\mathrm{gw}\le1.7\times10^{-7}$ by that time~\cite{Pagano:2015hma}.

The frequencies $f=0.85\,\upmu\mathrm{Hz}$ and $f=0.15\,\mathrm{mHz}$ mentioned above correspond to the $n=2$ harmonics of the Earth-Moon and Earth-LAGEOS systems, respectively.
The corresponding forecast sensitivity curves are strongly peaked in both cases, since the coupling to the $n=2$ harmonic is by far the strongest for low-eccentricity orbits like that of the Moon ($e\approx0.055$) and LAGEOS ($e\approx0.0045$)~\cite{Blas:2021mpc}.
The next most sensitive frequency in both cases is the $n=1$ harmonic, which is sensitive to $\Omega_\mathrm{gw}\ge3.2\times10^{-4}$ for LLR and $\Omega_\mathrm{gw}\ge2.2\times10^{-2}$ for SLR at present, improving to $\Omega_\mathrm{gw}\ge2.5\times10^{-7}$ and $\Omega_\mathrm{gw}\ge7.5\times10^{-5}$ respectively by 2038.
(See Fig.~1 in the Supplemental Material for the individual sensitivities of each harmonic of the Earth-Moon system.)

While binary pulsars are not able to compete with the laser-ranging experiments in terms of sheer sensitivity, their forecasts cover a much wider frequency band, spanning nearly five decades in frequency from $\approx6\,\mathrm{nHz}$ up to $\approx0.2\,\mathrm{mHz}$.
This is partly due to the range of orbital periods of various systems, and partly to the large eccentricities of many of these binaries, which gives them sensitivity to much higher harmonics.
The overall binary pulsar sensitivity curves shown in Fig.~\ref{fig:all-constraints} are computed by combining the overlapping PI curves of 215 binaries from the \href{https://www.atnf.csiro.au/research/pulsar/psrcat/}{ATNF pulsar catalogue}~\cite{Manchester:2004bp}.
The most stringent forecast sensitivity from this combined curve is $\Omega_\mathrm{gw}\ge8.2\times10^{-4}$ at $f=14$--$25\,\mathrm{nHz}$ with present data, expected to reach $\Omega_\mathrm{gw}\ge7.5\times10^{-7}$ by 2038.

Fig.~\ref{fig:all-constraints} also shows various potential SGWB signals around the $\upmu$Hz band probed by our proposed binary resonance searches.
The most important to mention here are the phase transition spectra, partly because FOPTs are a robust prediction of many well-motivated extensions to the Standard Model of particle physics~\cite{Caprini:2015zlo,Caprini:2018mtu,Caprini:2019egz}, and partly because the spectral shape of a FOPT signal highlights the constraining power of binary resonance searches~\footnote{%
    Here we focus on the SGWB signal due to sound waves in the plasma, as this is expected to be the dominant contribution for most FOPTs~\cite{Caprini:2019egz}.}.
While binary resonance probes are not competitive with GW interferometers and PTAs in searching for SGWB spectra which are roughly flat over many decades in frequency (e.g. GWs from inflation or cosmic strings), they can prove extremely useful for spectra that are confined to a narrow frequency band.
FOPTs are a leading example of such a signal, producing a narrow spectrum with a peak frequency~\cite{Caprini:2015zlo}
    \begin{equation}
    \label{eq:fopt-peak-freq}
        f_*\approx19\,\upmu\mathrm{Hz}\times\frac{T_*}{100\,\mathrm{GeV}}\frac{\beta/H_*}{v_w}\qty(\frac{g_*}{106.75})^{1/6},
    \end{equation}
    and a peak intensity of
    \begin{align}
    \begin{split}
    \label{eq:fopt-peak-intensity}
        \Omega_\mathrm{gw}(f_*)\approx5.7\times10^{-6}&\times\frac{v_w}{\beta/H_*}\qty(\frac{\kappa\alpha}{1+\alpha})^2\qty(\frac{g_*}{106.75})^{-1/3}\\
        &\times\qty[1-\qty(1+2\tau_\mathrm{sw}H_*)^{-1/2}].
    \end{split}
    \end{align}
Here $T_*$ is the temperature at which the FOPT occurs, $\alpha$ is the energy density released by the FOPT in units of the radiation density at the transition epoch, $\beta$ is the inverse duration of the transition, $H_*$ is the Hubble rate at the epoch of the transition, $v_w$ is the bubble wall velocity, $\kappa$ is an efficiency parameter determined by $\alpha$ and $v_w$, and $g_*$ is the number of relativistic degrees of freedom in the plasma, which we normalise to the Standard Model value, $g_*^{(\mathrm{SM})}=106.75$.
The second line of Eq.~\eqref{eq:fopt-peak-intensity} is a suppression factor due to the finite lifetime of the sound waves, $\tau_\mathrm{sw}$, which is a function of $\alpha$, $\beta$, and $v_w$~\cite{Caprini:2019egz}.

In Fig.~\ref{fig:fopt} we perform a scan over the FOPT parameters $(T_*,\alpha,\beta/H_*,v_w)$ for transitions occurring between $T_*=10^{-3}\,\mathrm{GeV}$ and $10^7\,\mathrm{GeV}$, identifying regions of parameter space where the corresponding SGWB signal is expected to be detected by binary resonance searches and other GW probes by 2038.
We find that LLR and SLR are able to probe significant regions of the FOPT parameter space at $T_*\sim\mathrm{GeV}$ and $\sim100\,\mathrm{GeV}$ respectively.
While SLR is less sensitive than LISA and will provide only complementary information, LLR will probe a region of the parameter space that is not accessible by any other planned GW experiment, thus providing a unique and valuable contribution to the search for phase transitions in the early Universe.
FOPTs are only one example of a strongly-peaked SGWB spectrum, but they demonstrate that binary resonance searches (and LLR in particular) have unique GW discovery potential.

Another potential SGWB signal shown in Fig.~\ref{fig:all-constraints} is the stochastic common process identified by the NANOGrav collaboration in their 12.5-year PTA dataset~\cite{Arzoumanian:2020vkk}.
While there is not yet sufficient evidence for quadrupolar cross-pulsar correlations to confidently interpret this signal as being due to GWs, the values inferred for its amplitude and spectral tilt are consistent with those expected for the SGWB from a population of inspiralling supermassive binary black holes~\cite{Middleton:2020asl} (SMBBHs), as well as with several more exotic interpretations~\cite{Ellis:2020ena,Blasi:2020mfx,Vaskonen:2020lbd,DeLuca:2020agl,Buchmuller:2020lbh,Ratzinger:2020koh,Vagnozzi:2020gtf,Neronov:2020qrl,Kuroyanagi:2020sfw}.
Assuming that the spectrum seen by NANOGrav can be extrapolated into the $\upmu$Hz band, we find that present-day LLR data are able to probe some of the steeper spectra allowed by the NANOGrav data (roughly $\Omega_\mathrm{gw}\sim f^{1.8}$), which could correspond to a strongly blue-tilted~\footnote{Such spectra can avoid the existing LVK and $N_\mathrm{eff}$ constraints if one allows for a nonstandard thermal history~\cite{Kuroyanagi:2020sfw}.} inflationary tensor spectrum~\cite{Vagnozzi:2020gtf,Kuroyanagi:2020sfw}.
If instead we assume that the NANOGrav signal follows the $\Omega_\mathrm{gw}\sim f^{2/3}$ scaling expected from inspiralling SMBBHs, we find that the spectrum should be detectable with 2038 LLR data.
This provides further motivation for the binary resonance searches we propose, showing that LLR can probe the nature of GW signals detected in the nHz band by NANOGrav and other PTAs.

%%%%%%%%%%%%%%%%%%%%%%%%%%%%%%%%%%%%%%%%%%%%%%%%%%%%%%%%%%%%%%%%%%%%%%%%%%%%%%%
\emph{Summary and outlook.}---In this Letter we have demonstrated the potential for binary resonance searches to bridge the $\upmu$Hz gap in the SGWB spectrum, showing that high-precision data from pulsar timing and laser-ranging experiments may lead to the first discovery of (or stringent constraints on) the SGWB in this region.
In particular, the sensitive frequency band of LLR sits almost exactly halfway between those of LISA and PTAs, and is thus highly complementary to these experiments.

As an illustrative example of the constraining power of binary resonance searches, we have considered potential SGWB spectra from FOPTs, showing that near-future LLR and SLR data will be sensitive to a broad range of FOPT models, and that LLR in particular can probe regions of the FOPT parameter space that are inaccessible to all other GW experiments.
We have also shown that current and future LLR data can provide complementary information about nHz GW signals probed by PTAs, such as the candidate SGWB signal recently announced by the NANOGrav Collaboration.

Our results provide strong motivation for further work in this direction.
On the theory side, there is plenty of scope to extend our formalism, either to other gravitationally-bound systems (e.g. hierarchical triples, globular clusters) or other GW signal morphologies (e.g. transient and/or narrowband signals, even if not exactly on-resonance).
Ultimately, the most pressing future work is to develop SGWB search pipelines based on our code \href{https://github.com/alex-c-jenkins/gw-resonance}{\texttt{gwresonance}}, allowing us to efficiently study the $\upmu$Hz--mHz band, perhaps even to discover GW signals waiting for us in this as-yet-unexplored regime.
The history of both electromagnetic and GW astronomy gives us plenty of reasons to be optimistic about the outcomes of these searches, and their potential for scientific discovery.

%%%%%%%%%%%%%%%%%%%%%%%%%%%%%%%%%%%%%%%%%%%%%%%%%%%%%%%%%%%%%%%%%%%%%%%%%%%%%%%
\begin{acknowledgments}
    We thank Vitor Cardoso, Jordi Miralda-Escud\'e, James Millen, Joe Romano, and two anonymous referees for valuable feedback on this work.
    We are grateful to Richard Brito for sharing with us the SGWB spectra from ultralight bosons shown in Fig.~\ref{fig:all-constraints}, and to Marek Lewicki for providing us with the AION-km PI curve and enlightening us about FOPTs.
    We acknowledge the use of \href{https://numpy.org/}{\texttt{NumPy}}~\cite{Harris:2020arr} and \href{https://www.scipy.org/}{\texttt{SciPy}}~\cite{Virtanen:2020sci} in our Python code, as well as the MCMC sampler \href{https://emcee.readthedocs.io/en/stable/}{\texttt{emcee}}~\cite{ForemanMackey:2012ig} in producing the FOPT exclusion regions.
    Fig.~\ref{fig:all-constraints}, and Figs.~1--3 in the Supplemental Material, were produced using \href{https://matplotlib.org/}{\texttt{Matplotlib}}~\cite{Hunter:2007}, while Fig.~\ref{fig:fopt} was produced using \href{https://corner.readthedocs.io/en/latest/}{\texttt{corner.py}}~\cite{ForemanMackey:2016cor}.
    A.C.J. was supported by King's College London through a Graduate Teaching Scholarship.
    D.B. is supported by a ``Ayuda Beatriz Galindo Senior'' from the Spanish ``Ministerio de Universidades'', grant BG20/00228.
    D.B. acknowledges support from the Fundaci\'on Jesus Serra and the Instituto de Astrof\'isica de Canarias under the Visiting Researcher Programme 2021 agreed between both institutions. D.B. also acknowledges the hospitality of the Theoretical Physics Department of Universidad de Zaragoza.
\end{acknowledgments}

\bibliography{binary-resonance-short}
\clearpage
%%%%%%%%%%%%%%%%%%%%%%%%%%%%%%%%%%%%%%%%%%%%%%%%%%%%%%%%%%%%%%%%%%%%%%%%%%%%%%%

\appendix
\section{SUPPLEMENTAL MATERIAL}

\begin{figure*}[t!]
    \includegraphics[width=0.50\textwidth]{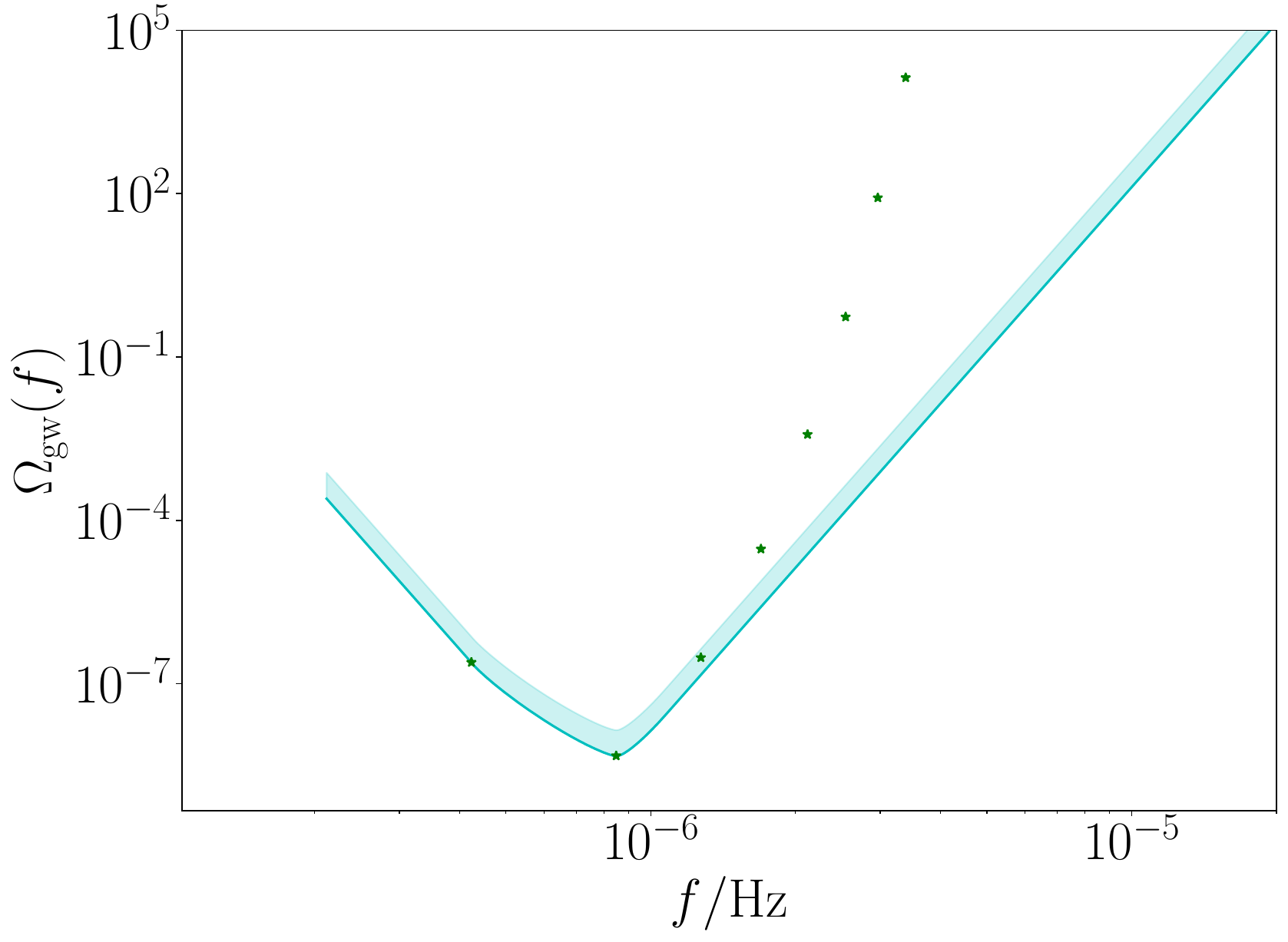}
    \includegraphics[width=0.49\textwidth]{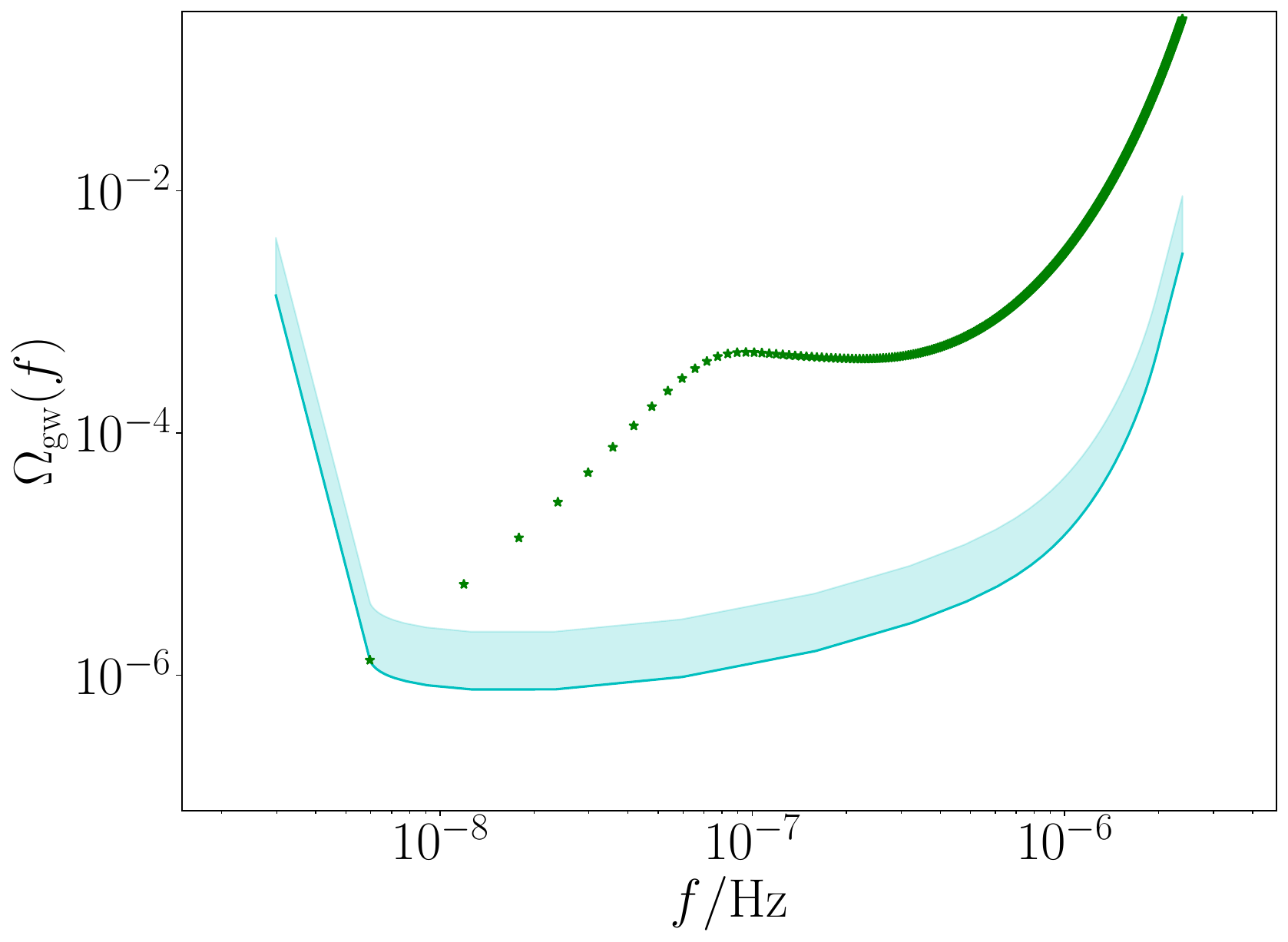}
    \caption{%
    Comparison of the continuous PI curves (cyan) and discrete frequency ``comb'' forecasts (green points) for two binary systems at 2038 sensitivity: the Earth-Moon system in the left panel, and the binary pulsar J1638-4725 in the right panel.
    }
    \label{fig:combs}
\end{figure*}

%-----------------------------------------------------------------------------%
\emph{Integrating the Fokker-Planck equation.}---On observational timescales, the evolution of the DF from ``sharp'' initial conditions can be well-approximated by considering just the first two moments of the Fokker-Planck equation (FPE)~\cite{Blas:2021mpc}; i.e., the mean vector $\bar{X}_i$ and covariance matrix $C_{ij}$ of the orbital elements.
The stochastic evolution of the mean vector due to binary resonance is typically much slower than the evolution due to deterministic effects, so it is convenient to separate the two effects by writing
    \begin{equation}
        \bar{X}_i(t)=\bar{X}_{0,i}(t)+\updelta\bar{X}_i(t),
    \end{equation}
    where the first term is the deterministic value, and the second term is the mean effect of the GW perturbations.
To leading order in $\Omega_\mathrm{gw}$, the evolution equations are then given by
    \begin{align}
    \begin{split}
    \label{eq:evol}
        \dot{\bar{X}}_{0,i}&=V_i,\\
        \updelta\dot{\bar{X}}_i&\simeq D^{(1)}_i-V_i+\updelta\bar{X}_j\partial_jV_i+\frac{1}{2}C_{jk}\partial_j\partial_kV_i,\\
        \dot{C}_{ij}&\simeq2D^{(2)}_{ij}+C_{ik}\partial_kV_j+C_{jk}\partial_kV_i,
    \end{split}
    \end{align}
    where $\partial_i\equiv\pdv*{X_i}$, and summation over repeated indices is implied.
We include in $V_i$ the general-relativistic precession of $\omega$ and $\varepsilon$ at first post-Newtonian order (1PN)~\cite{Poisson:2014gr}, and the decay of $P$ and $e$ due to GW emission at 2.5PN~\cite{Peters:1963ux}, but any other perturbations to the binary orbit can be included, e.g. due to tidal dissipation, or higher-order PN corrections.
(Neglecting these additional effects has little impact on our results here, but may be important for more refined searches in the future.)
The drift vector $D^{(1)}_i$ and diffusion matrix $D^{(2)}_{ij}$ are derived in a companion paper~\cite{Blas:2021mpc}; note that these are secularly averaged over the binary orbit, so that the evolution equations~\eqref{eq:evol} are only valid on timescales longer than the period $P$.
We include the first 400 harmonics in all of our integrations, although the evolution is almost always dominated by the first three harmonics.

By writing the FPE in this form, we have replaced a six-dimensional, second-order partial differential equation with 33 coupled, one-dimensional, first-order ordinary differential equations (six each for the deterministic mean elements $\bar{X}_{0,i}$ and the perturbations $\updelta\bar{X}_i$, with the remaining 21 coming from the independent components of the $6\times6$ symmetric matrix $C_{ij}$).
We integrate these equations numerically using a fifth-order Runge-Kutta method, as implemented in the \texttt{scipy.integrate} library~\cite{Virtanen:2020sci}.

%-----------------------------------------------------------------------------%
\emph{Sensitivity forecasts.}---For each of our binary resonance probes (MSPs, LLR, SLR), we assume an observational campaign in which the data are divided into intervals much shorter than the total observing time, but much longer than the binary period~\footnote{%
    We choose these intervals such that each is a year long; in principle greater sensitivity could be achieved by using fewer and longer intervals, as the accuracy of each orbital determination increases as the cube of the observing time~\cite{Blas:2021mpc,Mashhoon:1981cos}, but we are somewhat conservative here, as in practice there may be practical difficulties associated with combining large amounts of data coherently over much longer timescales.
    Future studies with real data will be required to determine the optimal analysis strategy in this regard.}.
We use a Fisher-forecasting approach to estimate the accuracy with which the orbital elements can be measured in each data interval, as quantified by the \emph{Fisher matrix}
    \begin{equation}
    \label{eq:fisher}
        F_{ij}\equiv\frac{1}{\sigma^2}\sum_a\pdv{\mathcal{O}_a}{X_i}\pdv{\mathcal{O}_a}{X_j}.
    \end{equation}
Here $a$ labels the individual data points, $\mathcal{O}_a$ is the observed quantity (for MSPs, the integrated pulse time-of-arrival or ``ToA''; for LLR and SLR, the ``normal point'' ranging distance), and $\sigma$ is the rms uncertainty in this quantity.
We assume that the number of data points is sufficiently large and uniformly distributed that the sum in Eq.~\eqref{eq:fisher} can be replaced by an integral averaging over the orbit.
We compute the derivatives $\pdv*{\mathcal{O}_a}{X_i}$ with respect to the orbital elements analytically, using the Blandford-Teukolsky timing formula~\cite{Blandford:1976pt} for MSPs and Kepler's equations for the ranging distance.

Given the stochastic nature of the GW-induced variations in the orbital elements, one might worry that these variations could be degenerate with the intrinsic noise of the observations, and could therefore be absorbed into $\sigma$.
However, we can convince ourselves that this is not the case by considering the data residuals caused by the orbital evolution (including secular effects from GWs) relative to a model in which $\Omega_\mathrm{gw}=0$ (i.e., with fixed orbital elements).
These residuals will generically grow over time, for two reasons: first, that the perturbations to the orbital elements are themselves expected to grow over time (roughly like $\propto\sqrt{t}$ in most cases, as expected for a random walk), and second, that even a constant offset in the orbital elements would generally cause the residuals to grow over time, due to the accumulation of relative phase between the the orbital model and the true orbit.
As a result, this effect cannot be absorbed into the intrinsic noise $\sigma$ (which is assumed to be stationary), and should thus be detectable with sufficient data.
(One could also hope to obtain further evidence for the GW-driven nature of this orbital evolution by combining data from several binaries with overlapping GW frequency sensitivities, and checking that each system shows orbital evolution consistent with the same SGWB spectrum.)

We assume the SGWB search is carried out by performing a likelihood-ratio test, comparing the maximum log-likelihood of the observed set of orbital elements under the assumption of a power-law SGWB spectrum $\Omega_\mathrm{gw}\sim f^\alpha$ to their log-likelihood in the absence of GWs,
    \begin{equation}
        \Lambda(\vb*X)\equiv2\max_{\Omega_\mathrm{gw}}\ln\frac{p(\vb*X|\Omega_\mathrm{gw})}{p(\vb*X|0)}.
    \end{equation}
In the limit of many observation intervals, this statistic is asymptotically $\chi^2_1$-distributed due to Wilks' theorem~\cite{Casella:2002stat} (where $\chi^2_1$ denotes the chi-square distribution with one degree of freedom), such that an observed value of $\Lambda\ge3.841$ would correspond to a detection of the SGWB with $95\%$ confidence.

The expectation value of the likelihood-ratio statistic in the presence of a SGWB signal is
    \begin{equation}
        \ev{\Lambda}_{\Omega_\mathrm{gw}}=\sum_tF_{ij}\qty(C_{ij}+\updelta\bar{X}_i\updelta\bar{X_j})-\ln\det(\delta_{ij}+F_{ik}C_{kj}),
    \end{equation}
    and can be calculated for a given SGWB spectrum by integrating the FPE moment equations~\eqref{eq:evol} over the duration of the observing campaign.
We thus estimate the detection threshold for a given experiment and for a given SGWB power-law index $\alpha$ by finding the smallest SGWB amplitude such that $\ev{\Lambda}\ge3.841$, using a numerical root-finding procedure.
We then iterate this procedure over different power-law indices, $\alpha=-10,-9.75,-9.5,\ldots,+10$, and take the maximum value of the resulting set of power-law curves at each frequency to construct the PI curves~\cite{Thrane:2013oya} shown in Fig.~1 of the main text.
The resulting curves represent the SGWB sensitivity of the binary, under the assumption that the SGWB spectrum is reasonably well-modelled as a power law with $|\alpha|\le10$ in the sensitive frequency band.
Fig.~\ref{fig:combs} shows how the shape of the resulting PI curve depends on the ``comb'' of constraints at each of the binary's resonant frequencies.
For low-eccentricity cases such as the Earth-Moon system ($e\approx0.055$), the $n=2$ harmonic is by far the most sensitive, giving a PI curve which is sharply peaked at this frequency (left panel of Fig.~\ref{fig:combs}).
On the other hand, high-eccentricity systems such as the binary pulsar J1638-4725 ($e\approx0.955$) can have sensitivity out to harmonics of order $n\sim100$ or more, giving much broader PI curves (right panel of Fig.~\ref{fig:combs}).

%-----------------------------------------------------------------------------%
\emph{Binary pulsars.}---We extract the orbital elements of 322 binary MSPs from the \href{https://www.atnf.csiro.au/research/pulsar/psrcat/}{ATNF pulsar catalogue}~\cite{Manchester:2004bp}, discarding 106 due to incomplete information, as well as the extremely wide binary J2032+4127, whose 46~yr period~\cite{Lyne:2015oua} means that the system has completed less than one complete orbit~\footnote{%
    Our evolution equations~\eqref{eq:evol} describe the \emph{secular} evolution of the binary on timescales much longer than the orbital period.
    It would be interesting to extend our formalism to cover sub-orbital timescales, similar to the treatment of \citet{Desjacques:2020fdi} for binaries perturbed by an oscillating axion field~\cite{Armaleo:2020yml,Blas:2019hxz}; we leave this for future work.}
     since its discovery in 2009~\cite{Abdo:2010en}.
For the remaining 215 MSPs, we extract the period $P$, eccentricity $e$, and argument of pericentre $\omega$; for near-circular systems $e\le10^{-3}$ the latter two are replaced by the Laplace-Lagrange parameters $\zeta=e\sin\omega$, $\kappa=e\cos\omega$, as these are more numerically stable when $e$ is very small.
The strongest GW constraints typically come from binaries with longer periods, although the sensitivity also depends on the eccentricity and argument of pericentre in a more complicated way---see the companion paper for details~\cite{Blas:2021mpc}.

The inclinations of binary MSPs are generally poorly-determined due to a degeneracy with the (often unknown) masses of the pulsar and its companion.
For most of the 215 systems, we assume a pulsar mass of $m_p=1.35\,M_\odot$ and an inclination of $I=\uppi/3$, as this corresponds to the median value of the companion mass $m_c$, which we extract from the catalogue.
In order to refine our results, we replace these values with more accurate mass and inclination determinations from the literature for the following MSPs, which produce the best SGWB bounds from our sample: J0737-3039A~\cite{Kramer:2006nb} (the double pulsar), B1913+16~\cite{Weisberg:2016jye} (the Hulse-Taylor system), B2127+11C~\cite{Jacoby:2006dy}, B1534+12~\cite{Fonseca:2014qla}, J1829+2456~\cite{Haniewicz:2020jro}, B2303+46~\cite{vanKerkwijk:1999xj}, J0045-7319~\cite{Kaspi:1996pul}, J1903+0327~\cite{Freire:2010tf}, J1740-3052~\cite{Madsen:2012rs}, and B1259-63~\cite{Miller-Jones:2018waj}.

\begin{figure}[t!]
    \includegraphics[width=0.48\textwidth]{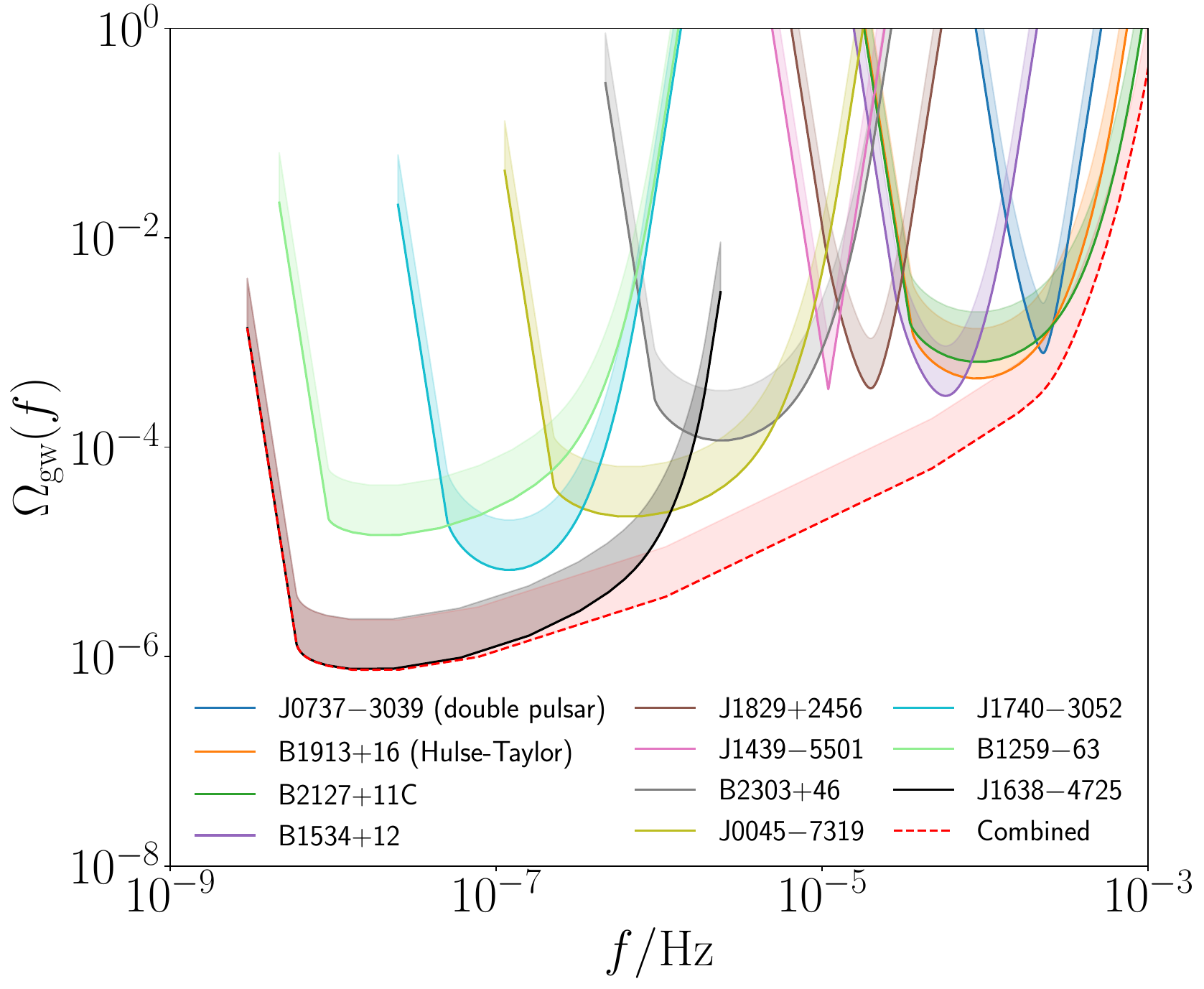}
    \caption{%
    SGWB PI curve forecasts from 11 binary pulsar systems at 2038 sensitivity.
    The red dashed curve shows the combined bound from these systems plus 204 others in the ATNF catalogue~\cite{Manchester:2004bp}, and corresponds to the red dashed curve in Fig.~1 of the main text.
    The assorted shapes of the curves shown here depend on the binary orbital parameters (particularly the eccentricity), and illustrate the utility of our formalism in accurately capturing the response of each system to the SGWB.}
    \label{fig:binary-pulsars-2038}
\end{figure}

Using these orbital elements and masses, we integrate the evolution equations~\eqref{eq:evol} from sharp initial conditions, with the initial time set to the year in which each system was discovered.
With these details specified, the SGWB sensitivity is then set by the number of ToAs per observing interval and the rms timing noise $\sigma$ associated with each ToA.
We assume each ToA corresponds to a 10-minute integration time.
For our 2021 sensitivity curves, we assume each system is monitored for two weeks every year, with ToAs being gathered for two hours every day within this period; this corresponds to the data cadence for B1913+16~\cite{Hui:2012yp}, and gives 168 ToAs per year.
We further assume $\sigma=1\,\upmu\mathrm{s}$.
For our 2038 sensitivity curves, we assume an observing campaign of 365 ToAs per year (i.e., 10 minutes of observations per pulsar per day) with $\sigma=80\,\mathrm{ns}$, which is the forecast 10-minute ToA uncertainty of next-generation radio telescopes like SKA~\cite{Liu:2011cka}.
We assume this observing campaign covers the entire period from 2021 to 2038, which is somewhat optimistic as SKA has not yet begun its pulsar timing observations; however, since the size of the perturbations to the orbital elements grows over time, our forecast constraints depend primarily on the timing precision at the end of the campaign, as well as the total observing time, rather than on the exact details of how the timing precision improves over time.
By combining the resulting individual PI curves for each of our 215 MSPs, we obtain the joint constraint curve shown in Fig.~\ref{fig:binary-pulsars-2038}.

It is important to note that our 2038 bounds are based only on known pulsars.
However, the SKA and other future radio telescopes are expected to discover large numbers of new pulsars~\cite{Janssen:2014dka}, some of which may be in binaries with orbits that are particularly sensitive probes of SGWB resonance.
We make no assumptions about these as-yet undiscovered pulsars, meaning that our 2038 bounds are conservative in this sense.

%-----------------------------------------------------------------------------%
\emph{Laser ranging experiments.}---For our LLR results we use the Lunar orbital elements, Lunar mass, and Earth mass tabulated in Murray and Dermott's \emph{Solar System Dynamics}~\cite{Murray:2000ssd}.
We base our 2021 sensitivity calculations on the APOLLO experiment, which has been observing since 2006, collecting roughly 260 ``normal point'' range measurements per year with a rms uncertainty of $\sigma\approx3\,\mathrm{mm}$~\cite{Murphy:2013qya}.
For our 2038 sensitivity curve, we assume an observation campaign which collects 1040 normal points per year (four times the current level) with an order-of-magnitude improvement in precision, $\sigma=0.3\,\mathrm{mm}$ (this would likely require the installation of new retroreflectors on the Lunar surface~\cite{Murphy:2013qya}, as the degradation of the existing reflectors is currently the main impediment to LLR sensitivity improvements).
We emphasise that including only the APOLLO experiment represents a conservative estimate of LLR sensitivity, as this excludes other experiments which have been collecting LLR data since 1969 (albeit with much less precision than the APOLLO data).

For our SLR results we focus on the LAGEOS-I satellite, with a start date of 1976, and using the satellite mass and orbital elements tabulated on the International Laser Ranging Service \href{https://ilrs.gsfc.nasa.gov/missions/satellite_missions/current_missions/lag1_general.html}{LAGEOS webpage}~\cite{ilrs:lageos}.
We assume that 50,000 normal points are collected per year for our 2021 sensitivity curve~\cite{ilrs:lageos}, rising to 200,000 per year by 2038 (again, a factor of four increase), and assume the same normal point uncertainties as for LLR in both cases.

We note that any futuristic GW mission in the solar system focusing on the band of interest here may face the challenge of modelling the gravity gradient noise from asteroids~\cite{Fedderke:2020yfy}, though the latter is several orders of magnitude too small to affect the forecasts we present in this work.

%-----------------------------------------------------------------------------%
\emph{Solar system bounds.}---All of the binary resonance searches discussed in the main text rely on precision measurements of orbital elements over observational timescales of years to decades.
However, our theoretical framework and our code \href{https://github.com/alex-c-jenkins/gw-resonance}{\texttt{gwresonance}} can also be used to study the SGWB-induced evolution of binaries on much longer timescales, e.g. the evolution of planetary orbits since the formation of the Solar System $\sim4.5\,\mathrm{Gyr}$ ago.
This amplifies the size of the effect we are interested in, as the deviations in the orbital elements typically grow like the square root of the elapsed time.
However, this also entails a loss of precision, as the initial values of the orbital elements are unknown.

\begin{figure}[t!]
    \includegraphics[width=0.48\textwidth]{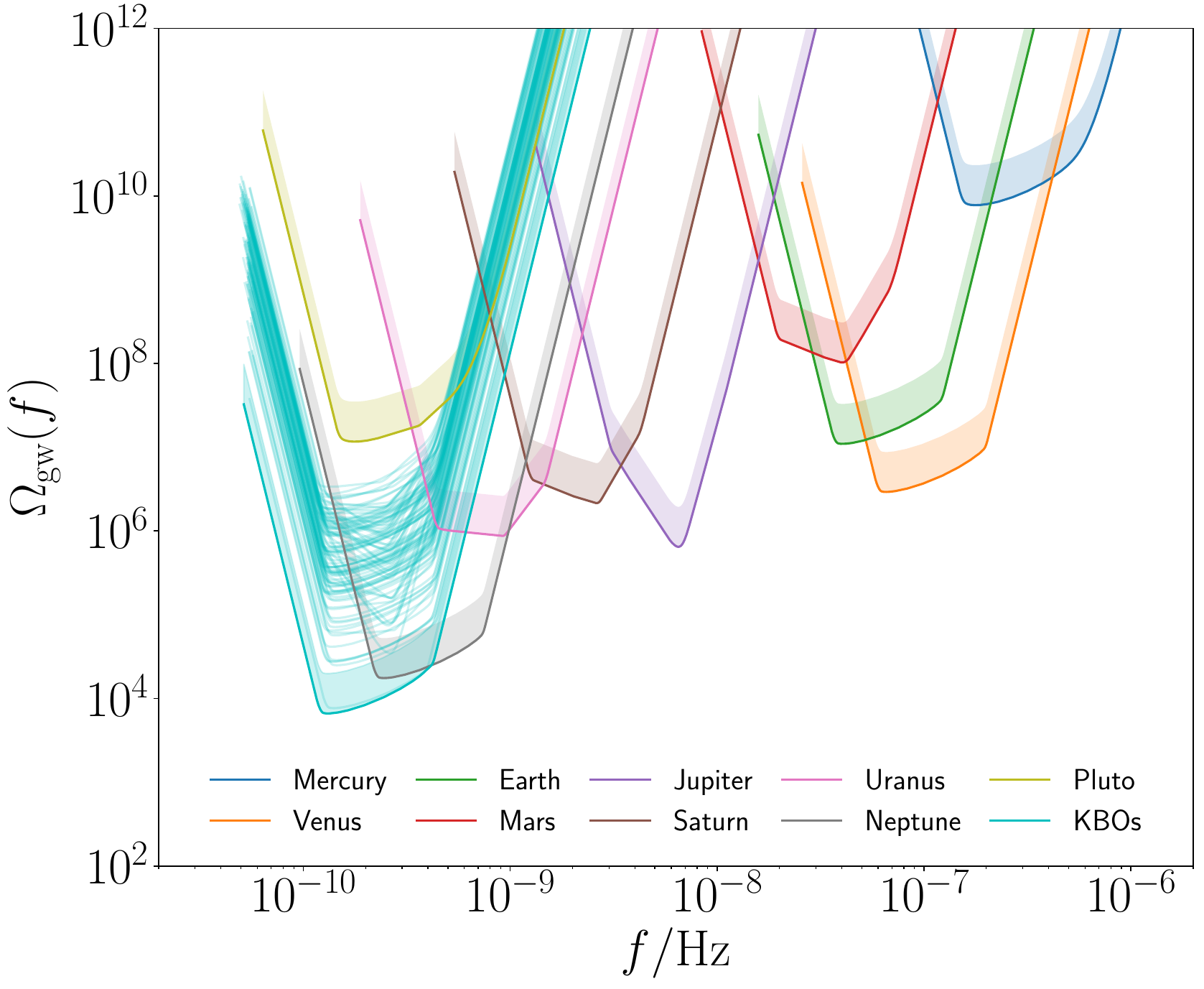}
    \caption{%
    SGWB PI curves inferred from the present-day orbital elements of various Solar System bodies.
    The faint cyan curves show constraints from 110 individual KBOs from the \href{https://ssd.jpl.nasa.gov/sbdb.cgi}{NASA/JPL Small-Body Database}, while the solid cyan curve shows the combined KBO constraint.
    }
    \label{fig:solar-system}
\end{figure}

In Fig.~\ref{fig:solar-system} we show SGWB constraints from the observed orbital elements of the eight Solar System planets, along with the dwarf planet Pluto and 110 classical Kuiper Belt Objects (KBOs).
We find that these are all orders of magnitude weaker than the precision binary resonance constraints possible with binary pulsars and laser ranging, with the strongest limit of $\Omega_\mathrm{gw}\le6.6\times10^3$ at $f=0.13\,\mathrm{nHz}$ coming from 523678 (2013 XB$_{26}$), a classical KBO on a very low-eccentricity orbit~\cite{jpl:sbdb}.

To produce these constraints, we integrate the evolution equations~\eqref{eq:evol} over the age of the solar system ($\sim4.5~\mathrm{Gyr}$), and comparing the present-day periods, eccentricities, and inclinations of various solar system bodies to the rms changes in each of these predicted due to binary resonance,
    \begin{equation}
    \label{eq:rms-x}
        \sigma_i=\sqrt{\updelta\bar{X}_i^2+C_{ii}},
    \end{equation}
    (no summation over the repeated index).
Since the SGWB tends to drive binaries towards longer periods, higher eccentricities, and larger inclinations, we can infer an upper limit on the SGWB intensity by requiring Eq.~\eqref{eq:rms-x} to be less than the present-day values of each of these quantities.
In doing so, we account for the redshifting of GWs over cosmological timescales, setting
    \begin{equation}
        \Omega_\mathrm{gw}=\Omega_{\mathrm{gw},0}\times(1+z)^4,\qquad f=f_0\times(1+z),
    \end{equation}
    with ``0'' subscripts denoting the present-day values that we place bounds on.
Since the solar system formed at redshift $z\approx0.41$, this can affect the final bounds by roughly a factor of $(1+z)^4\approx3.9$.
The redshifting of the GW frequency also broadens the shape of the resulting PI curve.

We extract the present-day orbital elements and masses of the eight planets and Pluto from Murray and Dermott~\cite{Murray:2000ssd}, as well as those of 110 dynamically cold ``classical'' KBOs from the \href{https://ssd.jpl.nasa.gov/sbdb.cgi}{NASA/JPL Small-Body Database}~\cite{jpl:sbdb}.
The individual PI curves of the KBOs are combined to give an overall PI curve for the Kuiper Belt constraint, which is dominated by 523678 (2013 XB26) at low frequencies, and by 79360 Sila-Nunam (1997 CS$_{29}$) at high frequencies, primarily due to their low eccentricities $e\approx0.007$.

%-----------------------------------------------------------------------------%
\emph{FOPT spectra.}---As mentioned in the main text, we include only the contribution from sound waves in the plasma, as this is generally expected to be the strongest component of the FOPT SGWB spectrum~\cite{Caprini:2015zlo}.
This contribution is given by~\cite{Schmitz:2020syl}
    \begin{equation}
        \Omega_\mathrm{gw}(f)=\Omega_\mathrm{gw}(f_*)\times(f/f_*)^3\qty[\frac{7}{4+3(f/f_*)^2}]^{7/2},
    \end{equation}
    where the peak frequency $f_*$ and peak intensity $\Omega_\mathrm{gw}(f_*)$ are given by Eqs.~(5) and~(6) of the main text respectively, subject to the requirement that the mean bubble separation,
    \begin{equation}
    \label{eq:bubble-separation}
        R_*=\frac{(8\uppi)^{1/3}}{\beta}\max(v_w,c_\mathrm{s}),
    \end{equation}
    is smaller than the Hubble scale $1/H_*$ (with $c_\mathrm{s}=1/\sqrt{3}$ the speed of sound in the plasma).
For the efficiency parameter $\kappa$ which appears in the peak intensity, we use the fitting functions in the appendix of \citet{Espinosa:2010hh} while for the sound wave lifetime we take~\cite{Ellis:2020awk}
    \begin{equation}
        \tau_\mathrm{sw}=R_*\times\qty(\frac{3\kappa}{4}\frac{\alpha}{1+\alpha})^{-1/2}.
    \end{equation}
In order to compute the $N_\mathrm{eff}$ constraints, we use the integrated form of this spectrum,
    \begin{equation}
        \int_{-\infty}^{+\infty}\dd{(\ln f)}\Omega_\mathrm{gw}(f)=\frac{343\sqrt{7/3}}{360}\Omega_\mathrm{gw}(f_*)\approx1.46\Omega_\mathrm{gw}(f_*).
    \end{equation}
(Strictly speaking this is an overestimate, as it includes frequencies $f\lesssim10^{-15}\,\mathrm{Hz}$ that do not contribute to $N_\mathrm{eff}$; however, this has negligible effect on the results in practice.)

We use the MCMC sampler \href{https://emcee.readthedocs.io/en/stable/}{\texttt{emcee}}~\cite{ForemanMackey:2012ig} to explore the FOPT parameter space, using the following priors:
    \begin{enumerate}
        \item transition temperature $T_*$: log-uniform in $[10^{-3},10^7]$ GeV;
        \item transition strength $\alpha$: log-uniform in $[10^{-3},10^3]$;
        \item inverse duration $\beta/H_*$: log-uniform in $[10^0,10^4]$;
        \item bubble wall velocity $v_w$: uniform in $[0.2,1]$.
    \end{enumerate}
We discard any samples for which the mean bubble separation~\eqref{eq:bubble-separation} is larger than the horizon, $R_*H_*>1$.
The resulting exclusion regions in Fig.~3 of the main text show FOPTs which can be detected at $\ge95\%$ confidence.

%%%%%%%%%%%%%%%%%%%%%%%%%%%%%%%%%%%%%%%%%%%%%%%%%%%%%%%%%%%%%%%%%%%%%%%%%%%%%%%
\end{document}